\pdfoutput=1
\documentclass[11pt,twoside,a4paper,cmspaper,final,collab]{cms-tdr}
\def\svnVersion{fe199c7}\def\svnDate{2026/04/13}\def\cmsCernNoTag{CERN-EP-2025-235}\def\cmsCernDate{\today}\def\cmsMessage{Published in Physical Review Letters as \href{http://dx.doi.org/10.1103/r53x-8fqp}{\doi{10.1103/r53x-8fqp}.}}
\begin{document}\cmsNoteHeader{EXO-24-031}

\newlength{\cmsFigWidth}

\ifthenelse{\boolean{cms@external}}{%
   \newcommand{\CP}{\ensuremath{C\!P}\xspace}
    \setlength{\cmsFigWidth}{\columnwidth}
    \renewcommand{\appendixname}{End Matter}
}{
  \newcommand{\CP}{\ensuremath{CP}\xspace}
   \setlength{\cmsFigWidth}{0.8\textwidth}
}
\newcommand{\PA}{{\HepParticle{A}{}{}}\xspace}

\cmsNoteHeader{EXO-24-031}

\title{Search for Light Pseudoscalar Bosons, Pair-Produced in Higgs Boson Decays in the Four-Electron Final State in Proton-Proton Collisions at \texorpdfstring{$\sqrt{s} = 13\TeV$}{sqrt(s) = 13 TeV}}

\date{\today}

\abstract{
A search for pairs of light neutral pseudoscalar bosons (\PA) resulting from the decay of a Higgs boson is performed. The search is conducted using LHC proton-proton collision data at $\sqrt{s} = 13\TeV$, collected with the CMS detector in 2016--2018 and corresponding to an integrated luminosity of 138\fbinv. The \PA boson decays into a highly collimated electron-positron pair. A novel multivariate algorithm using tracks and calorimeter information is developed to identify these distinctive signatures, and events are selected with two such merged electron-positron pairs. No significant excess above the standard model background predictions is observed. Upper limits on the branching fraction for $\PH\to\PA\PA\to4\Pe$ are set at 95\% confidence level, for masses between 10 and 100\MeV and proper decay lengths below 100\mum, reaching branching fraction sensitivities as low as $10^{-5}$. This is the first search for Higgs boson decays to four electrons via light pseudoscalars at the LHC. It significantly improves the experimental sensitivity to axionlike particles with masses below 100\MeV.
}

\hypersetup{
pdfauthor={CMS Collaboration},
pdftitle={Search for Light Pseudoscalar Bosons, Pair-Produced in Higgs Boson Decays in the Four-Electron Final State in Proton-Proton Collisions at \texorpdfstring{$\sqrt{s} = 13\TeV$}{sqrt(s) = 13 TeV}},
pdfsubject={CMS},
pdfkeywords={CMS, Higgs, axionlike particle}
}

\maketitle

The standard model (SM) of particle physics has been very successful at describing the fundamental particles and their interactions. However, many questions remain unanswered, such as the strong \CP problem~\cite{Wilczek:1977pj,Peccei:1977hh} and the nature of dark matter (DM)~\cite{Ramsey:1979bzw,Weinberg:1977ma,Moody:1984ba}. Particles and interactions beyond the SM are postulated to solve the above-mentioned limitations. For example, an elegant solution to the strong \CP problem is provided by the Peccei-Quinn mechanism~\cite{Peccei:1977hh,Kim:2008hd,Berezhiani:1989fp}, which introduces the axion. While the mass of the axion had originally been thought to be vanishingly small, well below 1\unit{eV}, recent studies indicate that mass values for the axion in the tens-of-\MeVns range are theoretically viable~\cite{Liu:2021wap,Girmohanta:2024nyf}. In certain DM models, the pseudoscalar axionlike particles (ALPs)~\cite{Peccei:1977hh,Jiang:2021dby,Kim:2008hd,Berezhiani:1989fp} are considered as mediators of the interaction between DM and SM particles, or even as direct candidates for DM particles, especially when they are light and weakly coupled to SM particles.

Consequently, searches for axions and light ALPs (collectively referred to as ALPs, hereafter) are of great interest and have been carried out extensively~\cite{Choi:2020rgn}. Furthermore, Ref.~\cite{Liu:2021wap} examines the implications of ALPs with a mass around 10\MeV, and hypothesizes an axion with a mass of 17\MeV from an observed excess in the electron-positron pair invariant mass spectrum, reported by the ATOMKI Collaboration~\cite{Krasznahorkay:2015iga} and recently examined by the MEG II~\cite{MEGII:2024urz} and PADME~\cite{Bossi:2025ptv} experiments. Finally, it is notable that the CERN LHC has a unique potential for the observation of ALPs, especially if the Higgs boson couples to them and decays into them~\cite{Bauer:2017nlg,Liu:2022tqn}.

For sub-\GeVns particles, electromagnetic decays can be used to both trigger and cleanly identify the events in LHC experiments. There are two possible electromagnetic decay channels for ALPs: electron-positron pairs and photon pairs. In the sub-\GeVns mass range, ALPs can be long lived because of their small mass and decay width. LHC searches from the CMS~\cite{CMS:2022fyt} and ATLAS~\cite{ATLAS:2023ian} Collaborations have set
the only limits on ALPs originating in Higgs boson decays with masses down to $\mathcal{O}(100)\MeV$. Searches for ALPs coming from Higgs boson decays, with masses as low as 10\MeV, or for ALPs decaying into electron-positron pairs, are unexplored. The reason is that the decay products of such low-mass ALPs are highly collimated for typical ALP momenta above 30\GeV. The search, therefore, needs innovative techniques to handle the merged electron-positron pairs, particularly in the case of ALP production via Higgs boson decays~\cite{Bauer:2017nlg,Liu:2021wap}.

In this Letter, a search is presented for ALPs with a mass in the range 10 to 100 MeV and a proper decay length ($c\tau$) below 100~\mum. The ALPs are produced in Higgs boson decays, and then each ALP decays into an electron-positron pair ($\PH\to\PA\PA\to4\Pe$).

The analysis strategy is to, first, reconstruct the ALP candidates from merged electron-positron pairs. Next, we reconstruct a Higgs boson candidate from a pair of ALP candidates and search for a localized excess in the four-electron invariant mass spectrum near the known Higgs boson mass. The signal of a Higgs boson decaying to ALPs would manifest as a narrow resonant peak in the four-electron invariant mass distribution on top of a smoothly falling and predominantly nonresonant background.

The study is based on data from proton-proton ($\Pp\Pp$) collisions at a center-of-mass energy $\sqrt{s} = 13\TeV$. The data were collected with the CMS detector~\cite{CMS:2008xjf,CMS:2023gfb} at the LHC from 2016 to 2018, corresponding to an integrated luminosity of 138\fbinv. Tabulated results are provided in the HEPData record for this analysis~\cite{hepdata}. The full likelihood for this analysis is provided in~\cite{cms_collaboration_2026_m332r-xsx83}.

The CMS apparatus~\cite{CMS:2008xjf,CMS:2023gfb} is a multipurpose, nearly hermetic detector, designed to trigger events of interest using a two-tiered trigger system~\cite{CMS:2020cmk,CMS:2016ngn} and identify electrons, muons, photons, and charged and neutral hadrons~\cite{CMS:2020uim,CMS:2018rym,CMS:2014pgm}. A global particle-flow (PF) algorithm~\cite{CMS:2017yfk} aims to reconstruct all individual particles in an event (PF candidates) by combining information from multiple subdetectors: the all-silicon inner tracker, the crystal electromagnetic calorimeter (ECAL), and the brass and scintillator hadron calorimeter (HCAL), all operating inside a 3.8\unit{T} superconducting solenoid, together with the gas-ionization muon system embedded in the flux-return yoke outside the solenoid. The PF candidates are then used to build \PGt leptons, jets, and missing transverse momentum~\cite{CMS:2018jrd,CMS:2016lmd,CMS:2019ctu}. Forward calorimeters, made of steel and quartz fibers, extend the pseudorapidity coverage provided by the barrel and endcap detectors. More detailed descriptions of the CMS detector, as well as the coordinate system and relevant kinematic variables, can be found in Refs.~\cite{CMS:2008xjf,CMS:2023gfb}.

In the Monte Carlo (MC) simulation, the signal is generated using Higgs boson production via the gluon-fusion process, modeled with the \POWHEG v2.0 generator at next-to-leading order (NLO) accuracy in quantum chromodynamics for the hard-scattering matrix element~\cite{Bagnaschi:2011tu,Nason:2004rx,Frixione:2007vw,Alioli:2010xd}. Parton showering, hadronization, and the Higgs boson decay are simulated with \PYTHIA 8.240~\cite{Sjostrand:2014zea}, including underlying event modeling with the CP5 tune~\cite{tuneCP5Paper}. The parton distribution function set is NNPDF3.1 at next-to-next-to-leading order (NNLO) accuracy~\cite{Ball:2017nwa}. Higgs boson-related backgrounds include gluon fusion production with decays into a pair of photons (FxFx prescription~\cite{Frederix:2012ps} for jet matching and merging) as well as the electron-positron-photon three-body Dalitz decay, modeled at NLO by \MGvATNLO v2.6.5~\cite{Alwall:2014hca}. The SM \PZ boson production is modeled at NNLO by the \POWHEG v2.0 generator with the \textsc{MiNNLO}$_{\textsc{PS}}$ prescription~\cite{Monni:2019whf,Monni:2020nks}. Double-photon production is modeled at NLO with \SHERPA 2.2.4~\cite{Bothmann:2019yzt} using the NNPDF3.0~\cite{NNPDF:2014otw} PDF, while single photon plus jet ($\PGg$+jets) is modeled with \PYTHIA 8.240.

Events are required to contain at least two reconstructed merged electron pairs (MEPs). Accordingly, high-level trigger paths that select two objects with significant electromagnetic energy~\cite{CMS-DP-2020-016} are used.

Despite the application of a novel end-to-end machine learning based reconstruction of the merged diphoton signature~\cite{CMS:2022wjj}, the diphoton search~\cite{CMS:2022fyt} was limited by the angular resolution of the CMS ECAL. In this search, we target ALPs decaying to electron-positron pairs. Unlike the diphoton channel, the two charged particles leave hits in the silicon tracker. The tracker has an angular resolution that is superior to the ECAL, which enables us to resolve highly collimated pairs and reach sensitivity to ALP masses as low as 10\MeV. However, the standard CMS reconstruction does not include a dedicated algorithm for highly collimated electron-positron pairs, and will typically reconstruct them as a single electron or misidentify them as converted photons. To address this, we have developed a set of dedicated reconstruction and identification techniques for merged electron-positron pairs. Henceforth, we refer to these as MEPs, since the electrons and positrons involved are not experimentally distinguished in this analysis. The MEPs are first reconstructed by matching two Gaussian-sum filter (GSF)~\cite{CMS:2020uim,Adam:2005bya} tracks from the silicon tracker to the same ECAL cluster; each track, extrapolated to the ECAL surface, must satisfy $\Delta R(\text{track},\text{ECAL cluster})<0.015$, where $\Delta R$ is the Euclidean distance in the $\eta$-$\phi$ plane. An ECAL cluster is a collection of energy deposits in the ECAL that contains the calorimeter shower from an electron or a photon~\cite{CMS:2020uim}. The GSF tracks are required to have a transverse momentum $\pt>5\GeV$ and fewer than two missing hits in the inner tracker. The associated ECAL cluster is required to have an energy larger than 25\GeV, which is above the thresholds for the triggers used in this analysis. The \pt of the reconstructed MEPs, obtained from the vector sum of the two GSF tracks, is further required to be consistent with the ECAL cluster \pt within 70\%. An adaptive vertex fitter~\cite{Fruhwirth:2007hz} is used to fit the vertex of the MEP using the two GSF tracks. The resulting $\chi^{2}$ value of the vertex fit is required to be smaller than five to reject failures of the vertex reconstruction. Successfully reconstructed MEPs are required to pass an isolation requirement, defined as the ratio of the summed transverse momentum and energy from the surrounding additional PF candidates within $0.015<\Delta R<0.3$. The isolation variable combines tracker, ECAL, and HCAL components, with the cutoff set at the 95\% signal efficiency, to reject jets misidentified as MEPs.

A gradient boosted decision tree (GBDT) classifier, implemented with the \textsc{XGBoost} package~\cite{10.1145/2939672.2939785}, has been developed to identify MEPs that pass the isolation criteria and to discriminate ALP-induced MEPs from background objects. The backgrounds mainly come from photon conversions, electrons with additional tracks falling within the $\Delta R<0.015$ cone centered on the electron momentum, and processes with jets that mimic the MEP signature. For example, processes such as \PZ boson, diphoton, and $\PGg$+jets production can contribute as backgrounds. The input variables include observables related to the MEPs, such as the shower shape variables from the ECAL, and tracking-related variables such as the angular separation between the two matched GSF tracks, as well as the displacement of the reconstructed vertex with respect to the primary vertex (PV). The PV is taken to be the vertex corresponding to the hardest scattering in the event, evaluated using tracking information alone, as described in Section 9.4.1 of Ref.~\cite{CMS-TDR-15-02}. We exclude input variables that are explicitly correlated with the energy scale of the ALP candidate, such as \pt and energy, to avoid biasing ALP identification to a specific kinematic region of the MEP. The model is then trained using samples with different ALP mass hypotheses. The identification model is trained separately for four different data-taking periods in 2016 (split into two subperiods because of a change in the detector conditions), 2017, and 2018. 

The identification model is trained using simulated samples of ALP signals, with various mass hypotheses, and simulated background samples, which include misreconstructed MEP candidates originating from photon conversions and misidentified electrons. The available samples are split into statistically independent subsets, with 80\% used for training and 20\% reserved for testing.

The GBDT working point (WP) chosen in the analysis, with a 70\% signal efficiency for the classifier, reduces about 75\% of MEP candidates from $\PZ\to\Pe\Pe$, about 55\% from $\PGg$+jets, and about 60\% from diphoton events, where the efficiencies are defined as the ratio between the number of MEPs passing the GBDT WP and the number of MEPs passing the isolation requirement. This WP optimizes the expected sensitivity reach and yields a sufficient number of events for the statistical analysis. The performance of this identification WP is further corrected with a $\PZ\to\PGm\PGm\PGg$ control sample, where a pure sample of converted photons is used as a proxy for MEPs. Data from different data-taking eras are separated to validate the MEP selection and extract scale factors (SFs), defined as corrections for efficiency differences in MEP identification between data and simulation. This is accomplished by fitting the \PZ boson peak in the three-body invariant mass distribution of the $\PGm\PGm\PGg$ system in both data and simulation. 

The $\PZ\to\PGm\PGm\PGg$ events used for this procedure are selected using standard CMS muon identification criteria~\cite{CMS:2018rym} together with the $\PZ$ boson mass constraint, ensuring a high-purity sample of photon conversions reconstructed as MEP candidates~\cite{CMS:2020uim}. The same selection is applied to simulated events to extract the efficiency correction factors.

The SFs are consistent with unity and are checked by varying the signal and background functions for the fit or the mass range of the fit. They are also checked in different bins of vertex transverse displacement with respect to the PV and the $\Delta R$ distance between the two GSF tracks, as motivated by the kinematic differences between the signal and background processes. The systematic shifts from the former are under 10\% for all cases, while for the latter, a closure within 5\% is found when measuring the SFs in different regions of the $\Delta R$ between the two GSF tracks. The efficiencies of the trigger paths are evaluated using $\PZ\to\Pe\Pe$ events that pass the MEP identification criteria used for the event selection. The single-leg efficiency is found to be 95\%, resulting in an overall event-level efficiency of about 90\%.

The Higgs boson candidates are reconstructed from the two successfully identified MEPs, with the reconstructed candidate mass $m_{4\Pe}$ found from the invariant mass of the system formed by the two MEPs, assumed to be massless. This approximation is justified since the ALP mass is negligible compared to the Higgs boson mass and detector resolution.

The MEP energy values are determined by the merged ECAL clusters. The observed yield of the signal, for the process $\PH\to\PA\PA\to4\Pe$, is then extracted from this mass spectrum by fitting the data points to a set of parametric shapes, for which different modeling methods are used for resonant and nonresonant processes. For example, the expected number of signal events after the full selection varies from five to ten, depending on the ALP mass and lifetime hypothesis, assuming a branching fraction of around $5\times10^{-5}$ corresponding to the observed sensitivity reach.

For the continuous background processes that originate from either prompt or nonprompt photon production and do not have a peaking structure around the Higgs boson mass, the falling shape of the mass spectrum is modeled from data. The background shape is determined by fitting the data in the sidebands of the 125\GeV Higgs boson mass peak, which are set as from 100\GeV to 180\GeV with the 115 to 135\GeV window of the Higgs boson mass excluded. A set of $F$~tests~\cite{fisher_chi2_1922} is used to constrain the model complexity and determine the nominal parameters for the final fit. The candidate background functions include polynomial functions, exponential functions, Bernstein polynomials, Laurent series, and power-law functions, which are widely used in the CMS $\PH\to\PGg\PGg$ analyses, such as in Refs.~\cite{CMS:2020cga,CMS:2024yhz}. The discrete profiling method~\cite{Dauncey:2015jinst_p04015} is introduced to account for the uncertainty of the parametric shape choices.

In addition to the nonresonant processes described above, which dominate the background, there is a nonnegligible resonant background. This is primarily from Higgs bosons decaying into two photons and the photons converting into electron-positron pairs, similar to ALPs. These converted photons are also reconstructed as MEPs and, therefore, produce a resonant background structure near the Higgs boson mass. This resonant background and the signal are modeled in approximately the same way. For both of them, a set of parametric shapes is used to model the peak structures from the intermediate Higgs boson resonance. The parametric shapes are fitted to simulated events that fulfill the event selection criteria. The parametric shape chosen for the peaking background is either a sum of multiple Gaussian functions, or a double-sided Crystal Ball function~\cite{Oreglia:1980cs,Gaiser:1982yw}, with the choice depending on the number of events, in order to ensure fit stability. While the former can be used with a sufficient number of events to exploit a more complicated signal shape behavior, the latter is used with a limited event count to ensure a single peak structure.

The dominant systematic uncertainties come from the SFs, affecting the identification and trigger efficiencies. The uncertainties from the SF fitting procedure and the phase space closure are already accounted for, as discussed above. The nonresonant background shape uncertainty is handled via the discrete profiling method discussed above. The systematic uncertainties on the shape of the distributions for the signal and resonant background processes arise from the ECAL cluster energy scale and resolution corrections, evaluated using varied simulation templates. The uncertainty on the integrated luminosity applies to the yield of simulated processes, ranging from 1.2\% to 2.5\% depending on the data-taking year, while the overall 2016--2018 uncertainty is 1.6\%~\cite{CMS:2016lumiPaper,CMSPASLUM17004,CMSPASLUM18002}. Finally, the theoretical uncertainty in the Higgs boson production cross section~\cite{deFlorian:2016spz} is included.

\begin{figure}[!htbp]
  \centering
    \includegraphics[width=\cmsFigWidth]{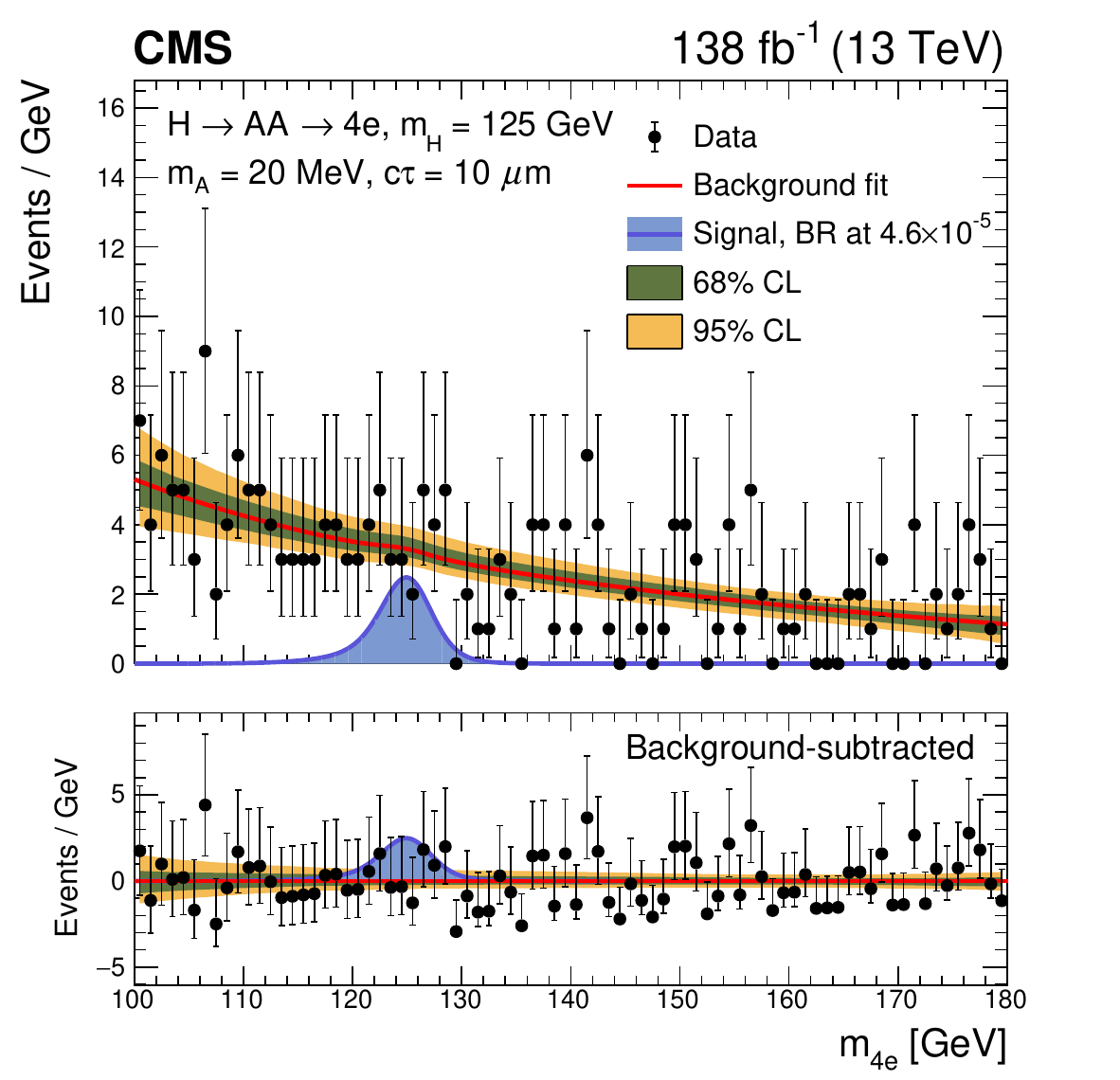}
  \caption{Invariant mass distribution of the four-electron system ($m_{4\Pe}$) for selected events (points), compared to the background-only fit (red) with its 68\% and 95\% \CL uncertainty bands (green and yellow). A nonstacked benchmark signal (blue) for a Higgs boson decaying to a pair of ALPs with $m_{\PA}=20\MeV$ and $c\tau=10\mum$ is overlaid and normalized to a branching ratio (BR) of $4.6\times10^{-5}$, which corresponds to the 95\% \CL upper limit value set by this analysis. The lower panel shows the same data after subtracting the background fit.}
  \label{fig:spb}
\end{figure}

The statistical analysis is performed with an unbinned profile-likelihood fit to the four-electron invariant mass spectrum using asymptotic formulas, implemented with the \textsc{Combine}~v8.2.0 package~\cite{CMS:2024onh}. The measured distribution of the four-electron invariant mass is compared to the signal-plus-background fit in Fig.~\ref{fig:spb}. No significant excess is observed over the background prediction. The observed limits are generally slightly higher than the expected limits, consistent with a small upward fluctuation of the data relative to the background-only prediction. The deviation remains within the 1-standard-deviation uncertainty band, and, therefore, is statistically compatible with the background-only hypothesis.

\begin{figure*}[!htbp]
  \centering
  \includegraphics[width=0.49\textwidth]{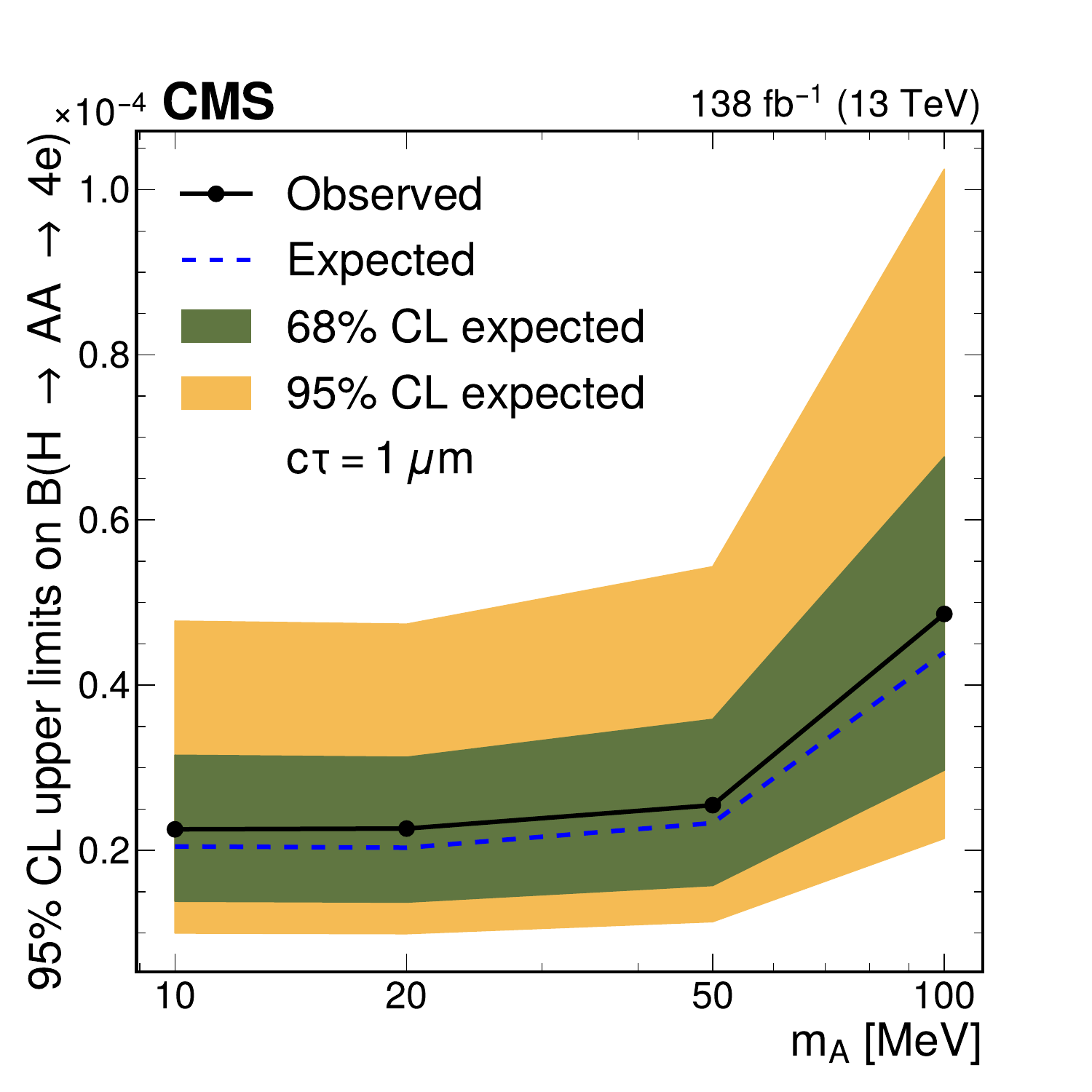} 
  \includegraphics[width=0.49\textwidth]{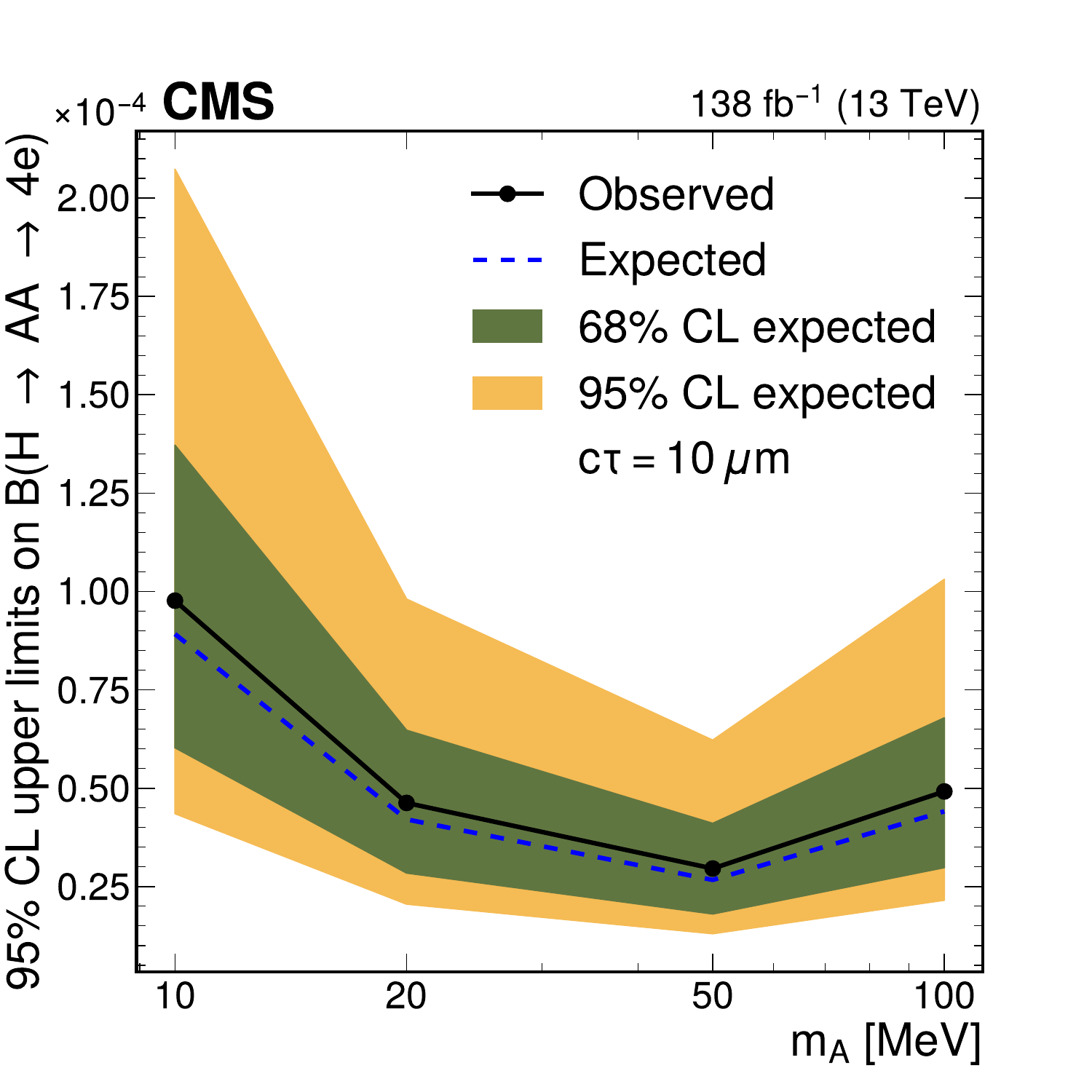} \\ 
  \includegraphics[width=0.49\textwidth]{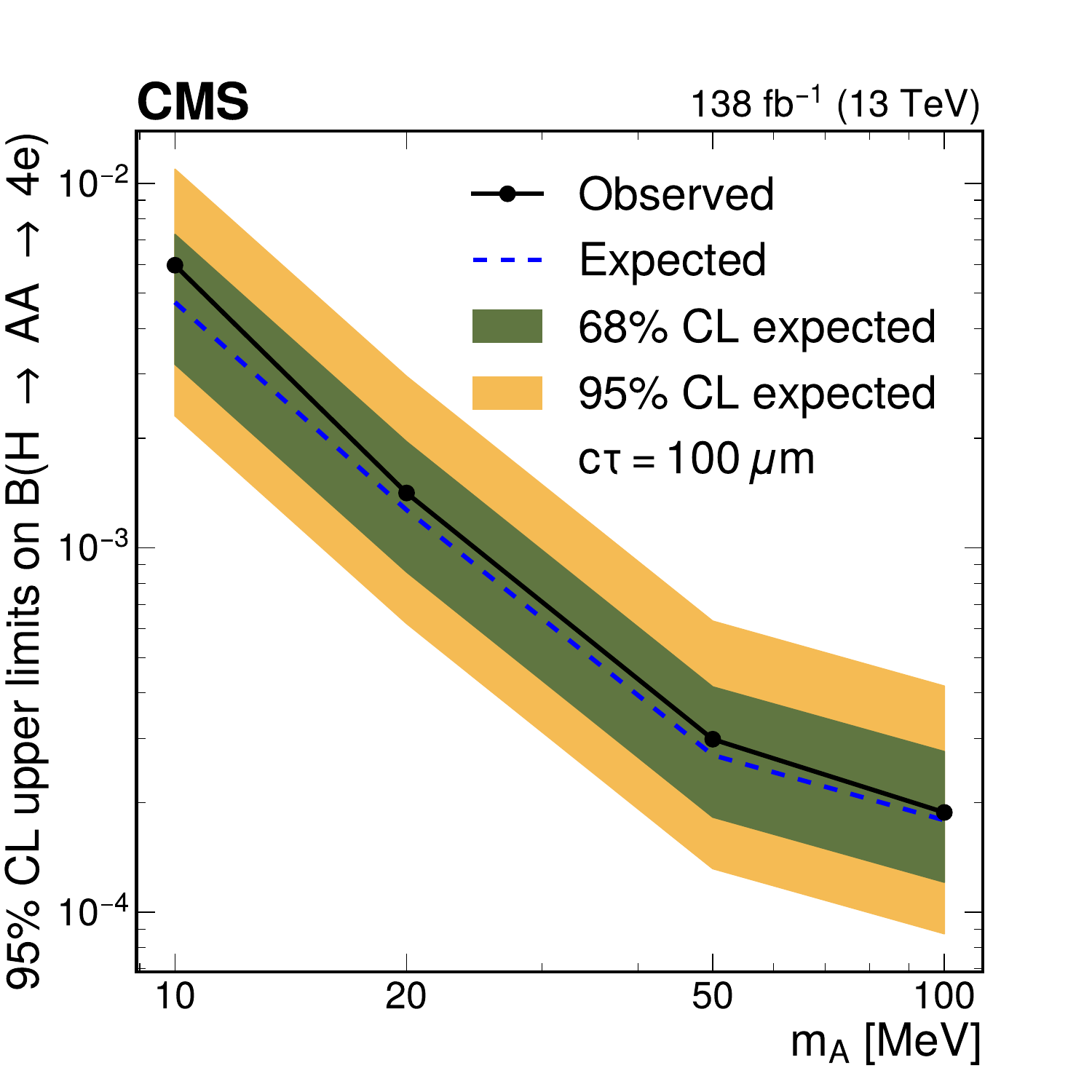}
  \includegraphics[width=0.49\textwidth]{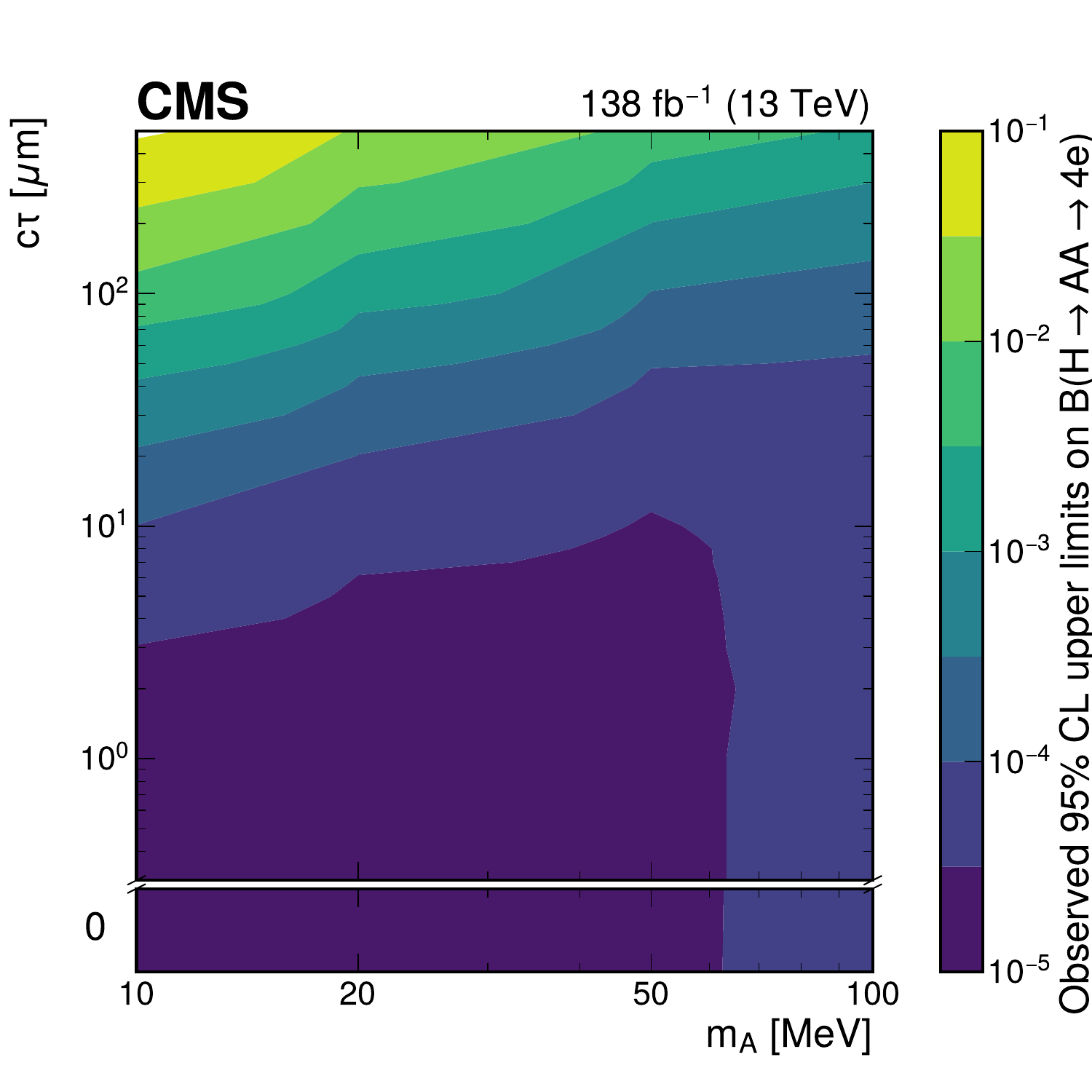}
  \caption{Observed (solid points) and expected (dashed lines) 95\% \CL upper limits on the Higgs boson branching fraction to a pair of ALPs decaying into electron-positron pairs ($\PH\to\PA\PA\to\Pe\Pe$), shown as a function of the ALP mass for benchmark proper decay lengths of 1\mum (upper left), 10\mum (upper right), and 100\mum (lower left). The green and yellow bands represent the 1 and 2 standard deviation confidence intervals around the expected limits. The lower right panel shows a map of the observed 95\% \CL upper limit, shown as a color scale, as a function of the ALP mass $m_{\PA}$, and proper decay length $c\tau$.}
  \label{fig:95cllimit}
\end{figure*}

Upper limits at 95\% confidence level (\CL) on the branching fraction for $\PH\to\PA\PA\to4\Pe$ are reported in Fig.~\ref{fig:95cllimit}. Figure~\ref{fig:95cllimit} shows the 95\% \CL upper branching fraction limits as functions of the ALP mass for $c\tau=1,10$, and 100\mum, as well as the map of these limits in the ALP mass versus lifetime plane. For the last plot, points with the additional lifetime hypotheses are created by reweighting the existing samples, using the exponential probability distribution of the decay probability. The upper limit on the decay branching fraction is observed to be as low as $\mathcal{O}(10^{-5})$ at 95\% \CL, in the mass range of 10 to 100\MeV and for proper decay lengths between 0 and 10\mum. The sensitivity degrades at larger masses because of wider opening angles of the electron pairs, and, also, at longer lifetimes because more ALPs start to fail the number of inner tracker hits requirement. An interpretation of these results in the effective ALP coupling model can be found in Appendix of the End Matter.

This analysis establishes the first direct limits on the Higgs boson exotic decay $\PH\to\PA\PA\to4\Pe$ for an axionlike particle \PA with a mass of $\mathcal{O}(10)\MeV$, reaching branching fraction sensitivities as low as $10^{-5}$. The limiting factor for the sensitivity of this analysis is the number of events in data, while the leading systematic uncertainty arises from the identification efficiency of merged electron-positron pairs. This search explores previously inaccessible parameter space and provides the most stringent constraints to date on this model. This search significantly improves the experimental sensitivity to axionlike particles with masses below 100~MeV, extending collider coverage to masses as low as 10~MeV for the first time. These results establish a new benchmark for future searches at the LHC.

\begin{acknowledgments}
We congratulate our colleagues in the CERN accelerator departments for the excellent performance of the LHC and thank the technical and administrative staffs at CERN and at other CMS institutes for their contributions to the success of the CMS effort. In addition, we gratefully acknowledge the computing centers and personnel of the Worldwide LHC Computing Grid and other centers for delivering so effectively the computing infrastructure essential to our analyses. Finally, we acknowledge the enduring support for the construction and operation of the LHC, the CMS detector, and the supporting computing infrastructure provided by the following funding agencies: SC (Armenia), BMBWF and FWF (Austria); FNRS and FWO (Belgium); CNPq, CAPES, FAPERJ, FAPERGS, and FAPESP (Brazil); MES and BNSF (Bulgaria); CERN; CAS, MoST, and NSFC (China); MINCIENCIAS (Colombia); MSES and CSF (Croatia); RIF (Cyprus); SENESCYT (Ecuador); ERC PRG, TARISTU24-TK10 and MoER TK202 (Estonia); Academy of Finland, MEC, and HIP (Finland); CEA and CNRS/IN2P3 (France); SRNSF (Georgia); BMFTR, DFG, and HGF (Germany); GSRI (Greece); NKFIH (Hungary); DAE and DST (India); IPM (Iran); SFI (Ireland); INFN (Italy); MSIT and NRF (Republic of Korea); MES (Latvia); LMTLT (Lithuania); MOE and UM (Malaysia); BUAP, CINVESTAV, CONACYT, LNS, SEP, and UASLP-FAI (Mexico); MOS (Montenegro); MBIE (New Zealand); PAEC (Pakistan); MES, NSC, and NAWA (Poland); FCT (Portugal);  MESTD (Serbia); MICIU/AEI and PCTI (Spain); MOSTR (Sri Lanka); Swiss Funding Agencies (Switzerland); MST (Taipei); MHESI (Thailand); TUBITAK and TENMAK (T\"{u}rkiye); NASU (Ukraine); STFC (United Kingdom); DOE and NSF (USA).
\end{acknowledgments}

\ifthenelse{\boolean{cms@external}}{\clearpage}{}
\bibliography{auto_generated}

\appendix

\ifthenelse{\boolean{cms@external}}{\section{Appendix: Interpretation of results in the effective ALP-Higgs coupling model}}{\section{Appendix: Interpretation of results in the effective ALP-Higgs coupling model}}
\label{endmatter}

To bridge the gap between the collider and intensity-frontier experiments, the reported limits are interpreted in an effective ALP coupling model, as described in Refs.~\cite{Bauer:2017nlg,Georgi:1986df}. In this framework, the ALP couples to SM fermions and the Higgs doublet through derivative interactions. The effective Lagrangian terms governing these interactions are described in Ref.~\cite{Bauer:2017nlg}, and the corresponding partial decay widths for the ALP decaying into fermion pairs and for the Higgs boson decaying into a pair of ALPs are, at LO, given by
  \begin{equation}
    \begin{aligned}
    \Gamma(\PA \to f\overline{f}) &=
      \frac{m_{\PA} m_{f}^{2}}{8 \pi \Lambda^{2}}
      c_{ff}^{2}
      \sqrt{1 - \frac{4 m_{f}^{2}}{m_{\PA}^{2}}}, \\
    \Gamma(\PH \to \PA\PA) &=
      \frac{v^{2} m_{\PH}^{3}}{32 \pi \Lambda^{4}}
      C_{\PA \PH}^{2}
      \left( 1 - \frac{2 m_{\PA}^{2}}{m_{\PH}^{2}} \right)^{2}
      \sqrt{1 - \frac{4 m_{\PA}^{2}}{m_{\PH}^{2}}}.
\label{eq:decay_width}
  \end{aligned}
\end{equation}
Here, $c_{ff}$ is the strength of the flavor-dependent coupling between the ALP and an SM fermion, $C_{\PA\PH}$ is the strength of the coupling between the ALP and the Higgs field, $\Lambda$ is the characteristic energy scale of such an effective theory (often denoted as $f_{\PA}$ in the literature on ALPs), $v$ is the vacuum expectation value of the Higgs field, and $m_{\PA}$, $m_{f}$, and $m_{\PH}$ denote the masses of the ALP, fermion, and Higgs boson, respectively.

\begin{figure}[!htp]
  \centering
    \includegraphics[width=\cmsFigWidth]{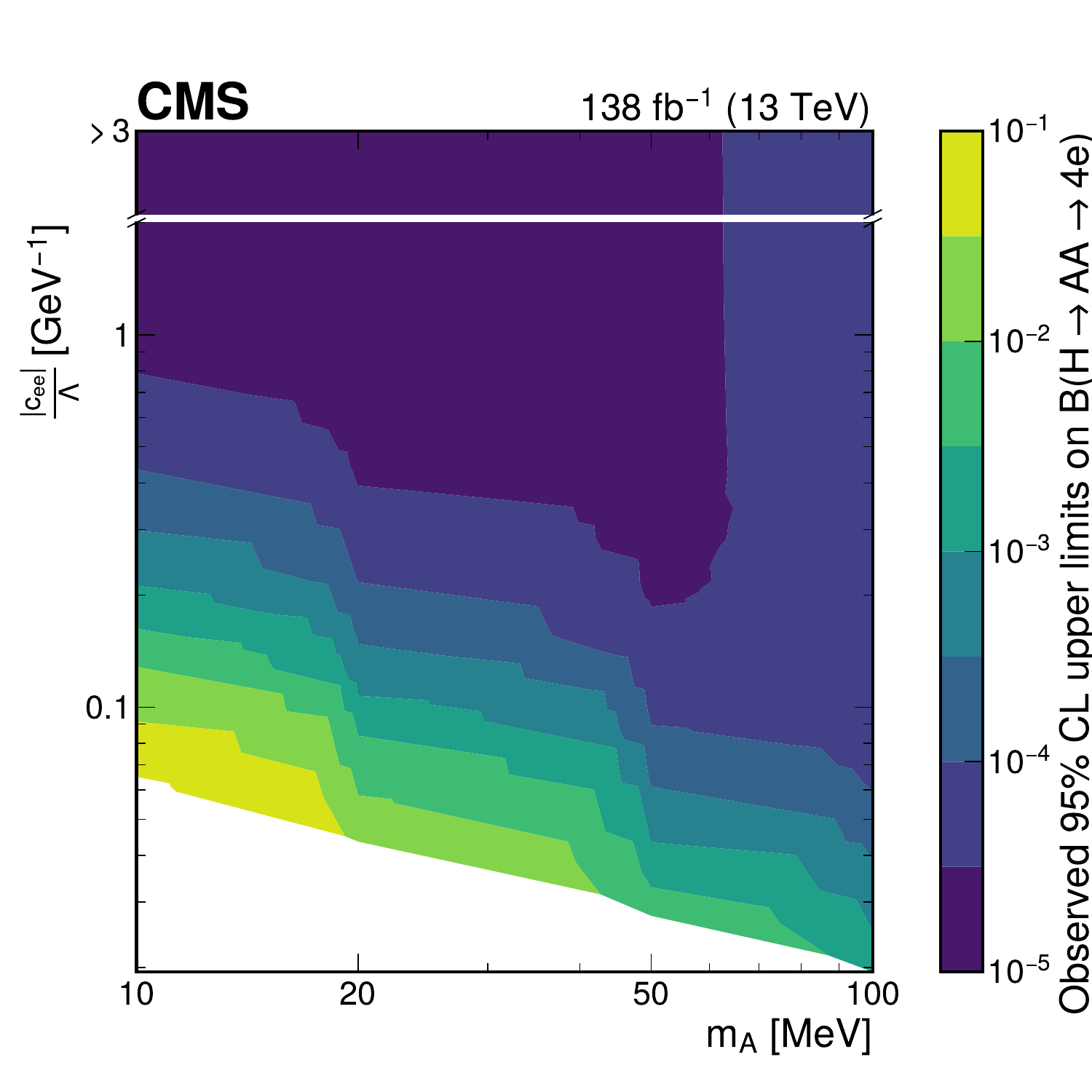}
  \caption{A map of the observed 95\% \CL upper limit on the Higgs boson branching fraction for $\PH\to\PA\PA\to4\Pe$, as a function of the ALP mass and the ratio of the ALP coupling to electrons to the energy scale of the ALP effective interaction.}
  \label{fig:reinterp_coupling}
\end{figure}

With the assumption that ALPs only decay to electron-positron pairs, the total decay width of the ALP is entirely determined by the partial width in this channel, thus, relating its lifetime to the parameter $c_{ff}/\Lambda$. Therefore, the results reported in Fig.~\ref{fig:95cllimit} can be reinterpreted as a function of this parameter, as shown in Fig.~\ref{fig:reinterp_coupling}.

\cleardoublepage \section{The CMS Collaboration \label{app:collab}}\begin{sloppypar}\hyphenpenalty=5000\widowpenalty=500\clubpenalty=5000\cmsinstitute{Yerevan Physics Institute, Yerevan, Armenia}
{\tolerance=6000
A.~Hayrapetyan, V.~Makarenko\cmsorcid{0000-0002-8406-8605}, A.~Tumasyan\cmsAuthorMark{1}\cmsorcid{0009-0000-0684-6742}
\par}
\cmsinstitute{Institut f\"{u}r Hochenergiephysik, Vienna, Austria}
{\tolerance=6000
W.~Adam\cmsorcid{0000-0001-9099-4341}, J.W.~Andrejkovic, L.~Benato\cmsorcid{0000-0001-5135-7489}, T.~Bergauer\cmsorcid{0000-0002-5786-0293}, M.~Dragicevic\cmsorcid{0000-0003-1967-6783}, C.~Giordano, P.S.~Hussain\cmsorcid{0000-0002-4825-5278}, M.~Jeitler\cmsAuthorMark{2}\cmsorcid{0000-0002-5141-9560}, N.~Krammer\cmsorcid{0000-0002-0548-0985}, A.~Li\cmsorcid{0000-0002-4547-116X}, D.~Liko\cmsorcid{0000-0002-3380-473X}, M.~Matthewman, I.~Mikulec\cmsorcid{0000-0003-0385-2746}, J.~Schieck\cmsAuthorMark{2}\cmsorcid{0000-0002-1058-8093}, R.~Sch\"{o}fbeck\cmsAuthorMark{2}\cmsorcid{0000-0002-2332-8784}, D.~Schwarz\cmsorcid{0000-0002-3821-7331}, M.~Shooshtari\cmsorcid{0009-0004-8882-4887}, M.~Sonawane\cmsorcid{0000-0003-0510-7010}, W.~Waltenberger\cmsorcid{0000-0002-6215-7228}, C.-E.~Wulz\cmsAuthorMark{2}\cmsorcid{0000-0001-9226-5812}
% ...
\par}
\vskip\cmsinstskip
\dag:~Deceased\\
$^{1}$Also at Yerevan State University, Yerevan, Armenia\\
$^{2}$Also at TU Wien, Vienna, Austria\\
% ...
\end{sloppypar}
\end{document}